\documentclass[aps,prl,twocolumn,superscriptaddress,groupedaddress]{revtex4}

\usepackage{graphicx}  
\usepackage{dcolumn}   
\usepackage{bm}        
\usepackage{amssymb}   

\def\be{\begin{equation}}
\def\ee{\end{equation}}
\def\bea{\begin{eqnarray}}
\def\eea{\end{eqnarray}}

\begin{document}

\title{Parametric Resonance of Entropy Perturbations in Massless Preheating}

\author{Hossein Bazrafshan Moghaddam,
Robert H. Brandenberger,
 Yi-Fu Cai, and Elisa G.M. Ferreira}

\email{bazrafshan, rhb, yifucai, elisafenu@physics.mcgill.ca}

\affiliation{Department of Physics, McGill University, Montreal,
QC H3A 2T8 Canada}



\begin{abstract}

Here, we revisit the question of possible preheating of entropy modes
in a two field model with a massless inflaton coupled to a matter
scalar field. Using a perturbative approximation to the covariant
method we demonstrate that there is indeed a parametric
instability of the entropy mode which then at second order
leads to exponential growth of the curvature fluctuation on super-Hubble
scale. Back-reaction effects shut off the induced curvature fluctuations,
but possibly not early enough to prevent phenomenological problems.
This confirms previous results obtained using different
methods and resolves a controversy in the literature.

\end{abstract}

\maketitle

\section{Introduction}

The ``reheating" phase is an important period of inflationary cosmology. It
describes the transfer of the energy density between the inflaton field, the
field which drives the inflationary expansion of space, and regular matter.
Without reheating, inflation would leave behind a universe devoid of matter,
not the kind of universe we observe. The energy transfer between the inflaton
field and regular matter is a consequence of couplings between the inflaton
field $\phi$ and ``regular matter". For simplicity, in studies of reheating
the regular matter which consists of fermions, gauge fields and scalars is
usually modelled by a scalar field $\chi$ which has some direct couplings
with $\phi$. The energy transfer at the end of the period of exponential
expansion of space from the inflaton field $\phi$ to regular matter $\chi$
was initially studied using first order perturbation theory \cite{DL, AFW, ASTW}.
However, such an analysis does not take into account the fact that
the inflaton field $\phi$ at the end of the period of inflation is in a highly
excited homogeneous condensate state and is not an assembly of free
perturbative quanta. A treatment of the energy transfer which takes
into account the coherence of the inflaton condensate was proposed in
\cite{TB, DK}. In these works, it was realized that the energy transfer
from $\phi$ and $\chi$ is driven by a parametric resonance instability.
This instability was worked out in detail in \cite{KLS1, STB, KLS2}, and the
word ``preheating" was coined to describe this energy transfer process.
The instability is exponential and typically leads to an energy transfer
which is rapid on the Hubble time scale $H^{-1}$, where $H$ is the Hubble
expansion rate at the end of the period of inflation. On the other hand,
the state of $\chi$ after preheating is not thermal, and to achieve a
thermal state of matter particles a second stage of reheating, namely
the thermalization stage is required. For recent reviews of reheating see
e.g. \cite{ABCM, Karouby}. We will not discuss the thermalization stage further since
we are interested in this paper in certain effects of the preheating phase.

From the mathematical point of view, the key aspect of preheating is that
the oscillations in time of the inflaton field $\phi$ after the end of
inflation lead to a periodically varying contribution to the mass term of
the $\chi$ field. The equation of motion for $\chi$ thus falls into the
category of those described by Floquet theory \cite{Floquet}, which
states that there are bands of Fourier modes of $\chi$ which experience
exponential growth. Since the inflaton field $\phi$ also couples to
gravity, the oscillations of $\phi$ lead to a periodically varying contribution
to the mass term in the equation of motion for cosmological perturbations,
as was first pointed out in \cite{NT1} and \cite{BKM}.
Hence, there is the possibility that
preheating can lead to a parametric instability in the amplitude of
cosmological perturbations, even on scales which are super-Hubble
at the end of inflation. If this were true, it would completely change the
usual predictions of inflationary models. In fact, the presence of a
resonant instability of cosmological fluctuation modes could lead to
an amplitude of fluctuations which is much larger than the observed
value, thus placing constraints on inflationary models.

In \cite{FB1} (see also \cite{Jedamzik, HK1, NT3}), it was shown that
in models with only adiabatic fluctuations there is no instability
of curvature fluctuations on super-Hubble scales. This is related
to the conservation of the comoving curvature fluctuation
variable $\zeta$ on super-Hubble scales \cite{AB, LV} (see also \cite{BA, BST, BK}
for early work). However, in certain two field models of inflation it was
argued in \cite{NT2, BV, FB2, Parry} that in the case of ``massless preheating"
(the inflaton having vanishing mass) there will be a preheating instability
for the entropy fluctuation mode, even on super-Hubble scales \footnote{In the
case of ``massive preheating" (the potential of the inflaton being dominated by
the mass term) there is no preheating of the entropy mode at linear order in
cosmological perturbation theory, as shown in \cite{HK2}.}. In fact,
it was shown in \cite{Zibin, Laurence} that back reaction effects will not be strong
enough to shut off the instability before the entropy mode has become
dominant. The presence of a preheating instability for the entropy mode
was confirmed in the analyses of \cite{HM} and \cite{TsBa}, extended to
the case of multifield generalized Einstein models in \cite{FFB}, and applied
to certain examples in \cite{examples}.

However, there remain concerns about the conclusions of \cite{FB1}. An
analysis using the ``separate universe" method \cite{TaBa} argues
that the preheating of entropy fluctuations is less effective. An analysis
using the $\delta N$ formalism, a method which is closely related to the
separate universe approach, finds that there is parametric resonance of
the entropy mode \cite{SY}, although in a subsequent paper \cite{SY2}
the same authors do not find substantial effects on the curvature fluctuations .
In addition, little amplification of the curvature fluctuations is observed
in the numerical work of \cite{Zhiqi}
which was based on a numerical implementation of the $\delta N$ formalism.

Since the entropy mode sees a growing curvature fluctuation on super-Hubble
scales, any parametric resonance instability of an entropy mode can lead
to an exponential growth of the curvature fluctuation during the preheating
stage. This is a potentially disastrous effect since the fluctuations could
well grow to become larger than the observed values. Hence, from the point
of view of inflation model building it is very important to determine whether
the parametric resonance instability of the entropy mode is robust. The goal
of this paper is to reconsider this question using different methods than
have been used before. Specifically we will make use of the ``covariant
formalism", a formalism developed in \cite{Langlois:2005qp}
(see also \cite{previous} for
earlier related work), a formalism which can be applied even non-perturbatively
(we, however, will use a perturbative truncation of the formalism).

Our study shows that the preheating of entropy modes is indeed
effective in the massless preheating toy model which we consider, and
that this leads to an exponentially growing contribution to the
curvature fluctuation. Since we have shown that our equations are
the perturbative limit of a consistent non-perturbative formalism,
we now have a better reason for arguing that the instability we find
will extend beyond the perturbative treatment.

Our goal is to demonstrate an instability of the model on cosmological
scales. Our methods obviously break down once the nonlinear regime
is reached. At this point, numerical methods used to study
nonlinear preheating effects (see e.g. \cite{Zhiqi2} for the first numerical
code for studying preheating which includes the metric fluctuations,
and \cite{Zhiqi, Easther, Frolov, Felder, Rajantie, Bellido, Tkachev})
would have to be applied.
These nonlinear effects, however, cannot reduce the amplitude of
curvature perturbations on cosmological scales, and hence we do
not consider them.

We emphasize that our results are not new. They confirm the conclusions
reached in earlier work (specifically in \cite{FB1}). However, our paper
resolves a controversy in the literature about the conclusions which were
reached in the earlier works.

\section{Massless Preheating}

In \cite{FB1} necessary conditions for the effectiveness of preheating
of the entropy mode of metric fluctuations have been discussed.
One of the conditions is that there is efficient parametric resonance
in the matter sector in the absence of gravitational fluctuations. A
model of massless inflation satisfies this condition. Hence, in this
paper we consider a toy model containing a massless inflaton field
$\phi$ coupled to a massive matter field $\chi$ with a potential
\begin{equation}
V(\phi, \chi) \,
= \, \frac{\lambda}{4}\phi^{4}+\frac{1}{2}g^{2}\phi^{2}\chi^{2} \, .
\end{equation}
The interaction term $\phi^{2}\chi^{2}$ allows for the decay of the
coherent inflaton configuration $\phi$ into massive $\chi$ excitations.

Up to the mass term for $\chi$ (which we will neglect in the following) our
model is conformally invariant.  Thus, via a conformal transformation we
can map our model into one living in Minkowski space-time, and
then study  preheating in Minkowski space-time. This is technically
much simpler than performing the calculation in the original variables in an
expanding universe. As pointed out first by \cite{Greene:1997fu},
the structure of resonance in the matter sector depends in a crucial way
on the relation between the coupling constants $\lambda$ and $g^{2}$.
Indeed the only parameter responsible for the structure of resonance is the
ratio $\frac{g^{2}}{\lambda}$.

We are interested in studying the evolution of the entropy perturbation in
this two field model. To do this we first need to study the evolution of
the quantum field $\chi$ in the background of the classical field $\phi$.
This will be done in the rest of this section.

We consider the case in which the zero mode of the $\chi$ field
is zero and therefore $\chi$ has no effect on the inflationary
dynamics in the absence of quantum fluctuations. However,
quantum fluctuation of this field are continuously excited. Since
the equation of motion for $\chi$ is linear in $\chi$ each
Fourier mode evolves separately. We are interested in modes
whose wavelength today corresponds to cosmological scales. The
wavelength is smaller than the Hubble radius early during
the inflationary phase, grows relative to the Hubble radius
as a consequence of the accelerated expansion of space during
the period of inflation, and exits the Hubble radius a number
of e-folding times before the end of inflation. At that point,
the oscillations of the quantum fluctuations freeze out and
the modes can be squeezed.

Later on we will have to take into account the fact that the
ensemble of large scale fluctuations of the $\chi$ field will
generate an effective $\chi$ background in which smaller scale
$\chi$ modes live. We will find this background by averaging
over the large-scale fluctuations.

\subsection{Evolution of the Inflaton $\phi$}

First we review the dynamics of $\phi$ after the end of the
period of inflation. During inflation the effective potential for the
inflaton is $\frac{\lambda}{4}\phi^{4}$ . Therefore the Klein-Gordon
for the classical inflaton field $\phi$ is
\begin{equation}\label{KGphi}
\ddot{\phi}+3H\dot{\phi}+\lambda\phi^{3}=0 \, ,
\end{equation}
where $H$ is the Hubble parameter and `` $\dot{}$ `` is the derivative
with respect to physical time. For further simplification we
work with conformal time $\eta$ defined via
\[
ad\eta=dt \, .
\]
Then as we mentioned before we do a conformal transformation
\[
a\phi=\varphi \, .
\]
If we rewrite equation (2), we find
\begin{equation} \label{varphieq}
\varphi''+\lambda\varphi^{3}-\frac{a''}{a}\varphi=0 \, ,
\end{equation}
where `` $'$ `` denotes the derivative with respect to conformal time.

After the end of the period of slow-roll inflation, the background
$\Phi$ field will start anharmonic oscillations about $\Phi = 0$
since it lives in a confining potential. It is well known that for
a quartic potential $\lambda\phi^{4}$ the time average of
the time-averaged energy momentum tensor is traceless and hence the
equation of state (averaged over an oscillation period) is the same
as for radiation. Hence $a(\eta)\sim\eta$, and therefore the last term
in the equation (\ref{varphieq}) vanishes. Thus we obtain
\begin{equation} \label{varphieq2}
\varphi''+\lambda\varphi^{3}=0
\end{equation}
which has periodic solutions. To find these solutions we introduce
the dimensionless conformal time
\[
x\equiv\sqrt{\lambda}\tilde{\varphi}\eta \, ,
\]
where $\tilde{\varphi}$ is the constant amplitude of the oscillations
of $\varphi=\tilde{\varphi}f(x)$. The solution of equation (\ref{varphieq2}) can be
written in terms of Jacobi elliptic functions:
\begin{equation}
\varphi=\tilde{\varphi}cn(x-x_{0},\frac{1}{\sqrt{2}}) \, .
\end{equation}
We can approximate the elliptic cosine function by the leading term in its series
expansion, $\cos(x)$, which is very good approximation as discussed
in \cite{Greene:1997fu}. Therefore the solution of the Klein-Gordon equation
for the $\varphi$ field is
\begin{equation}
\varphi=\tilde{\varphi}\cos(x) \, .
\end{equation}

Before moving on to next section, for further reference it is useful
to find the form of scale factor in this theory \cite{Greene:1997fu}.
At the end of inflation and beginning of preheating the
effective potential is $\lambda\phi^{4}/4$ and the homogeneous value
of $\chi$ field is zero. Thus the Friedman equation is
\begin{equation}\label{Friedman}
H^{2} \, = \, \frac{8\pi}{3M_{p}^{2}}(\frac{1}{2}\dot{\phi}^{2}+\frac{\lambda\phi^{4}}{4}) \, ,
\end{equation}
where $H$ is the Hubble parameter. When averaging over several oscillations
of the inflaton field while $\phi\ll M_{p}$ we then find
\begin{equation}
a(x) \, \sim \, \sqrt{\frac{2\pi}{3}}\frac{\tilde{\varphi}}{M_{p}}x \, .
\end{equation}

\subsection{Evolution of the Preheat Field $\chi$}\label{section:chi_evolution}

In this subsection we study the evolution of the linear
mode functions of the $\chi$ field, and use the results
to determine an effective background $\chi$ field which
a fixed Fourier mode of the fluctuations will feel. As
discussed e.g. in \cite{Laurence}, this background is
obtained by integrating over fluctuations of wavelength
larger than the one we are considering.

At the classical level, the homogeneous value of the $\chi$ field is zero.
The effective background which a mode with wavenumber $k$ will
feel is generated by the quantum fluctuations of larger
wavelengths which have exited the inflationary Hubble radius
earlier, have been squeezed and decohered and hence become
classical (see e.g. \cite{Martineau, Kiefer}).
To find this effective background, we must first
solve the equation for the quantum fluctuations of the $\chi$ field.
For simplicity we take the spatial sections to be flat.

As it is standard in the field, we use the formalism of quantum
field theory in curved space-time. Since we are considering
a free quantum field $\hat{\chi}$, we can expand the field
in Fourier modes, and each Fourier mode in creation and
annihilation operators $\hat{a}_{k}^{+}$ and $\hat{a}_{k}$,
respectively
\bea
\hat{\chi}(t,\vec{x}) \, = \, \frac{1}{(2\pi)^{3/2}}\int d^{3}k
&\bigl[&\hat{a}_{k}\chi_{k}(t)\exp(-ik.x) \\
&+&\hat{a}_{k}^{+}\chi_{k}^{*}(t)\exp(ik.x) \bigr] \, , \nonumber
\eea
where the mode functions $\chi_{k}$ satisfy the following Fourier space
Klein-Gordon equation
\begin{equation} \label{Xeqt}
\ddot{\chi}_{k}+3H\dot{\chi}_{k}+(\frac{k^{2}}{a^{2}}+g^{2}\phi^{2})\chi_{k} \, = \, 0 \, .
\end{equation}
In conformal time and considering
conformal transformation
\be
a\chi \equiv X
\ee
as well as a conformal transformation of the $\phi$ field
\be
a\phi \equiv \varphi
\ee
we find
\begin{equation} \label{Xeq0}
X''_{k}+[\frac{k^{2}}{\lambda\tilde{\varphi}^{2}}+\frac{g^{2}}{\lambda}f(x)-\frac{a''}{a}]X_{k} \, = \, 0 \, ,
\end{equation}
where `` $'$ `` denotes the derivative with respect to dimensionless
conformal time, and $f(x)$ is the periodic with amplitude $1$.
As mentioned before, the last term in this equation vanishes
during massless preheating and the equation becomes:
\begin{equation} \label{Xeq1}
X''_{k}+[\frac{k^{2}}{\lambda\tilde{\varphi}^{2}}+\frac{g^{2}}{\lambda}\cos^{2}(x)]X_{k} \, = \, 0 \, ,
\end{equation}
where we have also inserted the approximate form of $f(x)$.
This equation has the structure of a Mathieu equation. To make it clear we
rewrite equation (\ref{Xeq1}) as follows:
\begin{equation} \label{Xeq2}
X''_{k}+[(\frac{k^{2}}{\lambda\tilde{\varphi}^{2}}
+\frac{g^{2}}{2\lambda})+\frac{g^{2}}{2\lambda}\cos(2x)]X_{k} \, = \, 0 \, .
\end{equation}

As we mentioned in the Introduction we will consider values of the
coupling constant for which it is known that there is preheating
in the matter sector. Hence, we consider the case $\frac{g^{2}}{\lambda}\simeq2$
since in this case all long wavelength modes of the $\chi$ field are located in
the instability region of the Mathieu equation \cite{Greene:1997fu}. Therefore
the solution for $X_{k}$ will be of the form
\begin{equation}
X_{k}(x) \, = A_{1}\exp(\mu_k x)P_{1}(x,k) + A_{2}\exp(-\mu_k x)P_{2}(x,k) \, ,
\end{equation}
where $\mu_k$ is the so-called Floquet exponent \cite{Floquet}
which in this case has a positive real component (the Lyapunov
exponent),  and $P_{1}$ and $P_{2}$ are periodic functions of $x$ with
amplitude $1$ which appear in the solution of the Mathieu equation \cite{Floquet}.
Note that the period is determined by the period of the inflaton field,
and is independent of $k$.

Considering only the growing mode we need to determine the constant $A_{1}$
by fixing the initial conditions. Since preheating is preceded by a phase of
inflationary expansion, the initial conditions for preheating are determined
by the evolution of the field during inflation. This slow-roll inflation is
given by a quasi exponential expansion of the universe, where the Hubble
parameter is almost constant. During inflation, quantum $\chi$ field
perturbations (as well as $\phi$ perturbations) are created from vacuum
initial conditions on sub-Hubble scales. As the wavelengths of these fluctuations
are amplified in this phase relative to the Hubble radius, they eventually
exit the Hubble radius where they "freeze out" and may undergo squeezing.

To see whether squeezing occurs, we have to compare the magnitude of the
induced mass term in (\ref{Xeq0}), the term $\frac{g^{2}}{\lambda}f(x) X_k$,
with the squeezing term $\frac{a''}{a} X_k$. Thus, the condition for
squeezing is
\be \label{comp}
\frac{g^{2}}{\lambda} \, < \, \frac{a''}{a} \, .
\ee
Re-expressing the derivative with respect to the rescaled time
in terms of the regular time derivates, the condition (\ref{comp})
becomes (making use of $| \dot{H} | \ll H^{2}$)
\be \label{comp2}
H^2 \, > \, g^2 \Phi^2 \, .
\ee
Since during slow-roll it follows from (\ref{Friedman}) that
$H^2 \sim \Phi^4$, we see that (\ref{comp2}) will be more
easily satisfied for large values of the inflaton field.
Thus, to see if we get squeezing we need to determine the
range of values of $\Phi$ during slow-roll inflation.
Specifically, we need to determine the value of $\Phi$ at
the end of the slow-roll period.

During slow-roll, the second derivative term in (\ref{KGphi}) is
neglected and the kinetic term is negligible in the
Friedmann equation (\ref{Friedman}). Solving for the evolution
of $\Phi$ in the slow-roll approximation and using the
result to check when the kinetic contribution to $H$ ceases
to be sub-dominant yields the result
\be
\varphi _{end}^{2}\sim 2(6\pi G)^{-1}
\ee
for the value of $\varphi$ at the end of the slow-roll period.
We are interested whether there is squeezing for modes which exit
the Hubble radius a number $N$ Hubble times before the end
of inflation (for scales of cosmological interest we have $N \sim 50$).
Making a Taylor expansion in the evolution of $\varphi$ about the
endpoint of the slow-roll phase yields the lower bound
\be
\varphi(N) \, > \, \varphi_{end} \left( 1+N\frac{3}{2}\right)
\ee
for the value of $\varphi$ $N$ Hubble expansion times before the
end of inflation. Inserting this result into (\ref{comp2}), we can see that
the condition (\ref{comp2}) is satisfied provided $N > 2$.
This means that the quantum fluctuations of the entropy modes will be squeezed
between when they exit the Hubble radius and when $N = 2$, i.e.
essentially to the end of the slow-roll phase.

Since during the squeezing period $X_k \sim a$, the power spectrum of the
entropy modes from inflation becomes scale invariant and the initial conditions
for the $X$ modes at the beginning of the preheating phase are given by:
\begin{equation}
A_{1}(k)=\frac{1}{\sqrt{2k}}\frac{a(t_{R})}{a(t_{H}(k))}=H_I k^{-3/2}\,,
\end{equation}
where $t_{R}$ is the time when inflation ends. We will normalize the scale
factor such that $a(t_{R})=1$. Also, $t_{H}(k)$ is the time of horizon crossing,
with $a(t_{H(k)})=k/H_{I}$ and $H_{I}$ the Hubble rate during inflation. With that,
the solution for $X_{k}$ is:
\be \label{growing}
X_{k}(x) \, \simeq \, H_{I}k^{-3/2}\exp(\mu_k x)P_{1}(x,k) \, .
\ee
Thus we can observe clearly that the $X_{k}$ mode function is exponentially
growing,  which in turn leads to an exponentially growing number density
of $\chi$ particles. The value of the Floquet exponent
as a function of $k$ is shown in Fig. 1 \cite{Greene:1997fu}.
In this figure, the
vertical axis denotes the value of the Floquet exponent, the
horizontal axis labels $k$. Note that for the infrared modes
which we are interested in, the value of the Floquet exponent
is about $0.2$. Even though $\mu_k < 1$, the time scale of
the exponential instability is (while long
compared to the oscillation time) short compared
to the Hubble expansion time. Note that the analysis
has so far been in the absence of gravitational fluctuations.

\begin{widetext}

\begin{figure}[htbp]
\includegraphics[height=7cm]{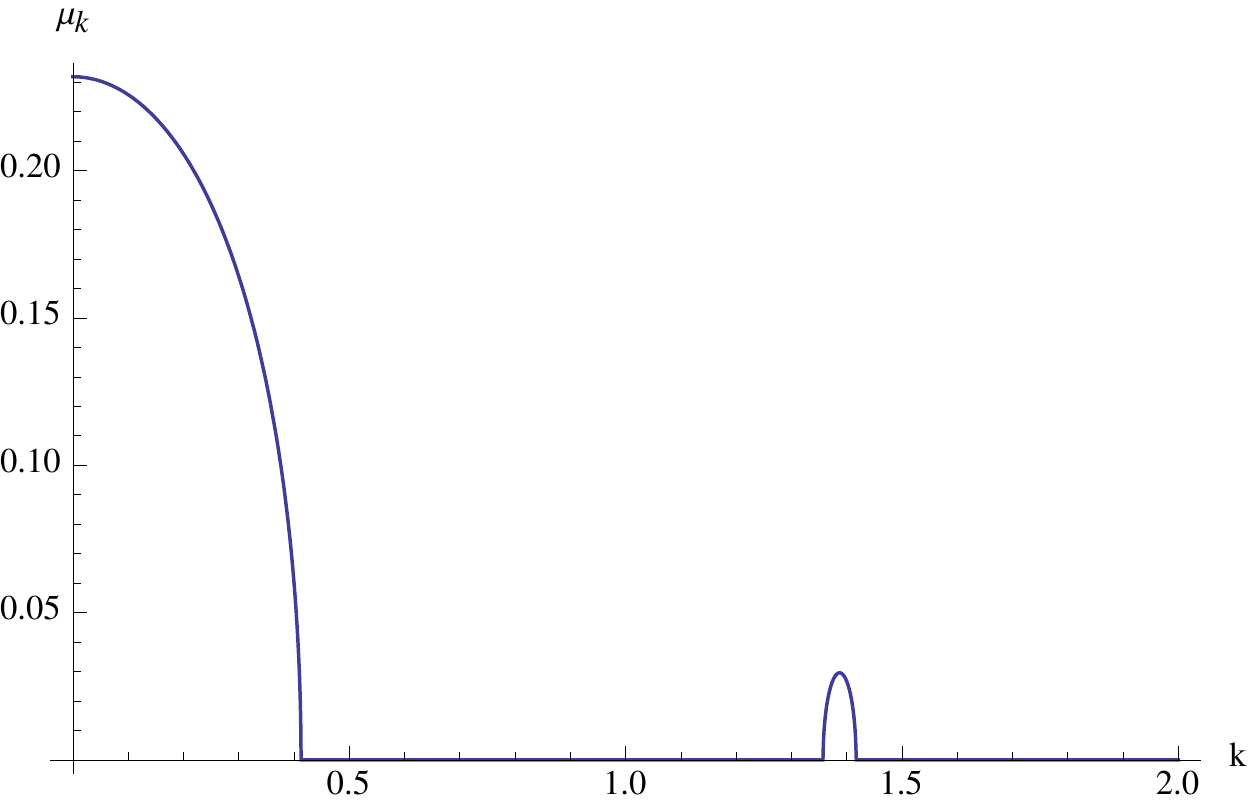}
\caption{Value of the Floquet exponent (vertical axis) as a function
of $k$ in units of $k^2 / (\lambda {\tilde{\phi}}^2)$.} \label{fig:1}
\end{figure}

\end{widetext}

The quantum fluctuations on large wavelength which were squeezed and
classicalized after exiting the Hubble radius will form
a background in which the metric fluctuations evolve. To find
the effective background value which a mode with wavenumber $k$
feels, we need to integrate over all wavelengths larger than $k^{-1}$
\footnote{This is a key point in our analysis. At strictly linear order
in perturbation theory the background value of $\chi$ would vanish.
However, in a particular patch of length $k^{-1}$ the average value of
$\chi$ will not vanish because of the presence of fluctuations on scales
which are in the infrared with respect to $k$. They will produce a
non-vanishing effective background. Our procedure corresponds to
integrating out infrared modes which are unobservable from the
point of view of the patch of interest.}
Thus, we define the effective background of the $\chi$ field on
a scale $k$ as
\begin{equation}
\chi_{\rm{eff}}(k) \equiv \bigl( \intop_{0}^{k}d^{3}k^{'}\, X_{k^{'}}^{2} \bigr)^{1/2} \, .
\end{equation}
To estimate this integral, recall that the period of $P(x, k)$ is independent
of $k$. To a good approximation we can take the Floquet exponent to
be independent of $k$ as long as the mode $k$ is in the infrared
instability band. Since we are interested in modes $k$ which are
cosmological today, the assumption that $k$ lies in this band will
be satisfied (recall that the infrared instability band runs from $k = 0$
to $k = k_c$, where $k_c$ is a microscopic scale). For vacuum
initial conditions the integral is dominated at  the upper end $k^{'} = k$,
and hence using (\ref{growing}) we obtain
\begin{equation}
\label{chi_eff}
\chi_{\rm{eff}}(k) \, \sim \, \sqrt{\pi}\, H \exp(\mu x)P_{1}(x,k) \, ,
\end{equation}
where we have dropped the index on the Floquet exponent.
The periodic function $P_{1}$ is \cite{Floquet}
a series of sine and cosine functions with different
coefficients and the leading term is $\sin(x)$ with unit coefficient
Therefore, the leading term for $P_{1}$ gives us
\begin{equation} \label{background}
\chi_{\rm{eff}}(k) \, \sim \, \sqrt{\pi}\, H \exp(\mu x)\sin(x)
\end{equation}
for the effective background value of the preheat field.

\section{Covariant Formalism}

In this section we review the covariant formalism following the
references \cite{Langlois:2005qp,Langlois:2006vv,Langlois:2008vk,Langlois:2010vx}
in which the formalism was developed (see also \cite{Rigopoulos} for related work).
The advantage of the covariant formalism is that it is based on variables which
vanish identically for the background cosmology. Hence the fluctuations
of these quantities are automatically gauge-invariant and can be used
for a non-perturbative analysis \cite{Stewart}.

In the following we first define the variables which are used in
the covariant analysis. Then we determine the dynamical
equations which they satisfy. We then make two simplifications.
First, we take the long wavelength approximation which is well
justified since we are interested in modes which are well outside
the Hubble radius during reheating. Finally, we solve the resulting
equations in the leading approximation.

\subsection{Case 1: A Perfect Fluid}

For instructive purposes we first review the covariant formalism in the
context of perfect fluid matter. We consider space-time as a manifold with a
preferred fluid flow direction which is characterized by a four-velocity $u^{a}$
which satisfies the normalization condition $u^{a}u_{a}=-1$.

The energy-momentum tensor for the perfect fluid is
\begin{equation}
T_{\; b}^{a} \, = \, (\rho+P)u^{a}u_{b} + Pg_{\; b}^{a}
\end{equation}
where $\rho$ and $P$ are energy density and pressure, respectively.
The spatial projection tensor orthogonal to the fluid four-velocity
is
\begin{equation}
h_{a\, b} \, \equiv \, g_{a\, b} + u_{a}u_{b}\, ,
\end{equation}
and satisfies the relations
\bea
h_{\; b}^{a}h_{\; c}^{b} \, &=& \, h_{\; c}^{a} \,\,\, {\rm{and}} \nonumber \\
h_{a}^{\; b}u_{b} \, &=& \, 0 \, .
\eea

The expansion parameter of space is given by
\begin{equation}
\Theta \, \equiv \, \nabla_{a}u^{a} \, ,
\end{equation}
where $\nabla_{a}$ is the covariant derivative. The
acceleration, $\dot{u^{a}}$ , is defined through the projected covariant
derivative along the four-velocity. To be more precise, let us define the
time evolution of any quantity in the covariant formalism by the
Lie derivative with respect to the flow direction in the manifold.
For a one form $Y_{a}$ the Lie derivative is defined by
\begin{equation}
\dot{Y}_{a} \, \equiv \, \mathcal{L}_{u}Y_{a} \, = \, u^{c}\nabla_{c}Y_{a} + Y_{c}\nabla_{a}u^{c} \, .
\end{equation}
For a scalar $f$ only the first term arises, i.e. $\dot{f} = u^{a}\nabla_{a}f$.
Therefore the acceleration is defined by
\begin{equation} \label{acc}
\dot{u^{a}} \, = \, \mathcal{L}_{u}u^{a} \, .
\end{equation}
For each comoving observer, we can define the logarithm
$\alpha$ of the local scale factor
by integrating $\Theta$ along the fluid world lines:
\begin{equation} \label{Theta}
\int d\tau\Theta \, \equiv \, 3\alpha \, .
\end{equation}

Key to the covariant approach is to make use of variables which vanish on the
unperturbed space-time.  Following \cite{Langlois:2005qp,previous,Bruni:1991kb}
one can define the ``projected covariant derivative'' operator
\be
D_{a} \, \equiv \, h_a^b \nabla_b \, .
\ee
It is the projection onto the hypersurface perpendicular to the vector field
tangent to the flow lines.  Next, one can introduce the
spatially projected covariant derivative (projected gradient) of the energy density
\begin{equation} \label{pertrho}
D_{a}\rho \, \equiv \, h_{a}^{\; b}\nabla_{b}\rho \, = \, \partial_{a}\rho + u_{a}\dot{\rho} \, ,
\end{equation}
of the pressure
\begin{equation}
D_{a}P \, \equiv \, h_{a}^{\; b}\nabla_{b}P \, = \, \partial_{a}P + u_{a}\dot{P} \,
\end{equation}
and of the expansion parameter
\begin{equation}
D_{a}\Theta \, \equiv \, h_{a}^{\; b}\nabla_{b}\Theta \, = \, \partial_{a}\Theta + u_{a}\dot{\Theta} \, .
\end{equation}
As these quantities vanish in FRW space-time, they yield a
fully geometrical and nonperturbative characterization of perturbations.

Knowing that, let us work out the generalized curvature and non-adiabatic
pressure perturbation in a geometrical way. Starting point are
the conservation equations for the energy-momentum tensor whose
first component fields the continuity equation
\begin{equation}
\dot{\rho} + \Theta(\rho+P) \, = \, 0 \, .
\end{equation}
Using the projected gradient of this equation one can define the curvature
covector as below \cite{Langlois:2006vv}
\begin{equation}
\zeta_{a} \, \equiv \, D_{a}\alpha - \frac{\dot{\alpha}}{\dot{\rho}}D_{a}\rho \, .
\end{equation}
Then the time evolution of this quantity is
\begin{equation} \label{zetaeq1}
\dot{\zeta}_{a} \, = \, \mathcal{L}{}_{u}\zeta_{a} \, = \, -\frac{\Theta}{3(\rho+P)}(D_{a}P-c_{s}^{2}D_{a}\rho)
\end{equation}
where $c_{s}^{2}\equiv\frac{\dot{P}}{\dot{\rho}}$ is the generalized
speed of sound as defined in \cite{Langlois:2006vv}. Comparing this
equation with the familiar equation of motion in linear theory where the
right-hand side of the equation is the non-adiabatic pressure perturbation,
we define non-adiabatic pressure covector as
\begin{equation}
\Gamma_{a} \, \equiv \, D_{a}P - \frac{\dot{P}}{\dot{\rho}}D_{a}\rho \, .
\end{equation}
Making use of the definitions in equations (\ref{acc},\ref{Theta},\ref{pertrho})
one can rewrite
the curvature and non-adiabatic pressure covectors in terms of ordinary gradients
\begin{equation}
\zeta_{a} \, = \, \partial_{a}\alpha - \frac{\dot{\alpha}}{\dot{\rho}}\partial\rho
\end{equation}
\begin{equation}
\Gamma_{a} \, = \, \partial P - \frac{\dot{P}}{\dot{\rho}}\partial\rho \, .
\end{equation}

For the case of a single scalar field one can show that non-adiabatic
pressure covector is
\be
\Gamma_{a} \, = \, 2\frac{\dot{\phi}}{\dot{\rho}}V_{,\phi}D_{a}\rho
\ee
which vanishes in the long wavelength approximation.
Then considering $\dot{\rho}=-\Theta\dot{\phi}^{2}$ in equation (\ref{zetaeq1}),
leads to the equation
\be \label{zetaeq2}
\dot{\zeta}_{a} \, = \, \frac{2}{3}\frac{V_{,\phi}}{\dot{\phi}^{3}}D_{a}\rho
\ee
for the time evolution of the curvature covector for the case of a single scalar
field. The right-hand side of this equation vanished in the long wavelength
approximation. This yields the conclusion that in the case of a single
perfect fluid the curvature fluctuation $\zeta$ is conserved on super-Hubble
scales at arbitrary order in perturbation theory.

\subsection{Case 2: A System of Two Coupled Scalar Fields}

The extension to the case of two scalar fields was given in \cite{Langlois:2006vv}.
The first step is to identify the adiabatic and the entropy
components of the fluctuations in this two field system. To do this we
use the formalism developed in  \cite{Gordon:2000hv} in which we are
given two scalar fields $\phi$ and $\chi$ which both have non-vanishing
backgrounds which are evolving in time.
The adiabatic field $\sigma$ is tangent to the field trajectory, the entropy
field $s$ is orthogonal to it. We can introduce the corresponding unit
vectors in two-dimensional field space via
\bea
e_{\sigma}^I \, &\equiv& \, \frac{1}{\sqrt{{\dot{\phi}}^2 + {\dot{\chi}}^2}}
\bigl( {\dot{\phi}}, {\dot{\chi}} \bigr) \, , \\
e_{s}^I \, &\equiv& \, \frac{1}{\sqrt{{\dot{\phi}}^2 + {\dot{\chi}}^2}}
\bigl(- {\dot{\chi}}, {\dot{\phi}} \bigr) \, ,
\eea
where $I$ is the field space index. The angle $\theta$ of the trajectory
in field space is then given (in the small angle approximation) by
\be
\theta \, = \, \frac{{\dot{\chi}}}{{\dot{\phi}}} \, .
\ee

Using the above definitions, we can set up the adiabatic and entropy field
covectors by taking the respective projective covariant derivatives of
the basis fields $\phi$ and $\chi$:
\bea
\sigma_{a} \, &\equiv& \, e_{\sigma}^{I}\nabla_{a}\varphi_{I} \,
= \, \cos\theta\nabla_{a}\phi + \sin\theta\nabla_{a}\chi , \\
s_{a} \, &\equiv& \,  e_{s}^{I}\nabla_{a}\varphi_{I} \,
= \, -\sin\theta\nabla_{a}\phi + \cos\theta\nabla_{a}\chi \, .
\eea
Note that $s_{a}$ is orthogonal to $u^{a}$ and we have $u^{a}s_{a} = 0$,
but this is not the case for adiabatic covector since
$u^{a}\sigma_{a} = \dot{\sigma}$.

The geometrical variables which describe the field perturbations are
obtained by taking the spatially projected version of the above equations
\bea
\sigma_{a}^{\bot} \, &\equiv& \,  e_{\sigma}^{I}D_{a}\varphi_{I} \,
= \, \sigma_{a} + \dot{\sigma}u_{a} \, , \\
s_{a}^{\bot} \, &\equiv& \,  e_{s}^{I}D_{a}\varphi_{I} \, = \, s_{a} \, .
\eea
Note that these fluctuations are well defined non-perturbatively.

From the Klein-Gordon equations for the $\phi$ and $\chi$ fields,
we can find the adiabatic and entropy components of the Klein-Gordon equations.
Using these equations we can find the evolution equation for the adiabatic
$\sigma_{a}$ and entropy $s_{a}$ covectors. The resulting equation
for the adiabatic component is \cite{Langlois:2006vv}
\bea
(\ddot{\sigma}_{a})^{\bot} &+& \Theta(\dot{\sigma})_{a}^{\bot} + \dot{\sigma}D_{a}\Theta
+ \bigl( V_{,\sigma\sigma} + \dot{\theta} \frac{V_{,s}}{\dot{\sigma}} \bigr)\sigma_{a}^{\bot} - D_{a} ( \nabla^c \sigma_c^{\bot} ) \nonumber \\
&=&
\bigl( \dot{\theta} - \frac{V_{,s}}{\dot{\sigma}} \bigr) s_{a})^{\centerdot}
+ \bigl( \ddot{\theta} - V_{,s\sigma} + \Theta \dot{\theta} \bigr) s_a - D_a Y_{(s)} \, ,
\eea
where
\be
Y_{(s)} \, = \, \frac{1}{\dot{\sigma}} \bigl( \dot{s}_a + \dot{\theta} \sigma_a^{\bot} \bigr) s^a \, .
\ee
The equation for the entropy component is \cite{Langlois:2006vv}
\bea
\ddot{s}_{a} &+& \bigl( \Theta - \frac{1}{\dot{\sigma}} ( \nabla^c \sigma_c^{\bot} - Y_{(s)} ) \bigr) \dot{s}_{a}
+ \bigl( V_{,ss} + \dot{\theta}^{2} - 2 \dot{\theta} \frac{V_{,s}}{\dot{\sigma}} \bigr) s_{a} \nonumber \\
&& - D_a ( \nabla_c s^c ) \\
&=& \frac{\dot{\theta}}{\dot{\sigma}} \bigl( D_a \Pi - \frac{\dot{\Pi}}{\dot{\sigma}} \sigma_a^{\bot}
- 2 \epsilon_a \bigr) - \frac{1}{\dot{\sigma}} \bigl( D_c s^c + Y_{(\sigma}) \bigr)^{\cdot} \sigma_a^{\bot}
\nonumber \\
&&+ D_a Y_{(\sigma)} \,
\nonumber
\eea
where $\epsilon_a$ is the covector associated with the comoving energy
density perturbation and
\be
Y_{(\sigma)} \, \equiv \, \frac{1}{\dot{\sigma}} \bigl( \dot{s}_a + \dot{\theta} \sigma_a^{\bot} \bigr) \sigma^{\bot, a} \, .
\ee

The first approximation we make is to linearize these equations
(the expansion parameter is the amplitude of the fluctuations, which in our
case is proportional to $\hbar$.) This yields
greatly simplified equations
\bea
(\ddot{\sigma}_{a})^{\bot} &+& 3 H (\dot{\sigma})_{a}^{\bot} + \dot{\sigma}D_{a}\Theta
+ \bigl( V_{,\sigma\sigma} - \dot{\theta}^2 \bigr)\sigma_{a}^{\bot} - D_{a} ( D^c \sigma_c^{\bot} ) \nonumber \\
&\simeq& \, 2 (\dot{\theta} s_a)^{\cdot} - 2 \frac{V_{,s}}{\dot{\sigma}} \dot{\theta} s_a \, ,
\eea
and
\bea
\ddot{s}_{a} &+& 3 H \dot{s}_{a}
+ \bigl( V_{,ss} +  3 \dot{\theta}^{2} \bigr) s_{a}  - D_a ( D_c s^c ) \nonumber \\
&\simeq& - 2 \frac{\dot{\theta}}{\dot{\sigma}} \epsilon_a \, .
\eea
The final approximation we make is to focus on long wavelengths, i.e. we work in the
leading order gradient expansion in which also the comoving energy density fluctuation
vanishes. This yields our final equations
\bea \label{adeq}
(\ddot{\sigma}_{a})^{\bot} &+& 3H (\dot{\sigma})_{a}^{\bot} + \dot{\sigma}D_{a}\Theta
+ \bigl( V_{,\sigma\sigma}  - \dot{\theta}^{2} \bigr) \sigma_{a}^{\bot} \nonumber \\
&\simeq& \,  2(\dot{\theta}s_{a})^{\centerdot} - 2\dot{\theta}\frac{V_{,\sigma}}{\dot{\sigma}}s_{a}
\eea
and
\begin{equation} \label{enteq}
\ddot{s}_{a} + 3 H \dot{s}_{a} +  \bigl( V_{,ss} + 3\dot{\theta}^{2} \bigr) s_{a} \, \simeq \, 0 \, .
\end{equation}
As is well known in the linear theory of cosmological perturbations (see e.g. \cite{MFB}
for an overview and \cite{RHBrev} for an introduction), the entropy fluctuations are not
affected by the amplitude of the adiabatic perturbation. On the other hand, entropy
fluctuations induce a growing adiabatic mode.

\section{Application of the Covariant Formalism to Massless Preheating}

Our goal in this section is to show that in massless preheating entropy fluctuations
are indeed parametrically amplified, and that this in turn leads to an exponentially
growing contribution to curvature fluctuations. As we shall see, the effect on the
curvature fluctuations is quadratic in the amplitude of the quantum fluctuations.
The entropy fluctuations themselves have an exponentially growing term which
is linear in the fluctuation amplitude. However, the coupling between the
entropy and the adiabatic mode is suppressed by an additional power which
comes from the fact that the background of the entropy field vanishes at
zero'th order.

For the case of massless preheating the equation for long wave linearized
entropy fluctuations (\ref{enteq}) becomes
\bea \label{enteq2}
\ddot{s}_{a} &+& 3 H \dot{s}_{a} +\Bigl[ ( 3\lambda\phi^{2} + g^{2}\chi^{2})\sin^{2}\theta
- 2g^{2}\phi\chi\sin2\theta \nonumber \\
&+& g^{2}\phi^{2}\cos^{2}\theta + 3\dot{\theta}^{2} \Bigr] s_{a} \, \simeq \, 0 \, .
\eea

To analyze this equation of motion we make a couple of approximations and use the
following setup:
\begin{itemize}
\item As mentioned before, at the beginning of preheating the overall homogeneous value
of the $\chi$ field is zero (the effective $\chi$ field on a scale $k$ will
be non-vanishing but of linear order in the amplitude of the fluctuations).
\item We use the relation for $\theta$ in the small angle approximation introduced
in the previous subsection:
\begin{equation}
\theta \, \equiv \, \frac{\dot{\chi}}{\dot{\phi}}
\end{equation}
for the instantaneous angle between the background trajectory and the $\phi$
field direction in field space. Therefore at the beginning of preheating
the angle is of linear order in the fluctuations.
\item We will use the result from \cite{Gordon:2000hv} for $\dot{\theta}$
in the large scale limit
\begin{equation}
\dot{\theta} \, = \, -\frac{V_{,s}}{\dot{\sigma}} \, ,
\end{equation}
where $V,s$ is the derivative of the potential with respect to the entropy
component. We use another result which is shown in \cite{Gordon:2000hv}
\be
V_{,s} \, = \, -V_{,\phi}\sin\theta + V_{,\chi}\cos\theta \, .
\ee
At the beginning of preheating we have $V_{,\chi} \simeq \chi$
which like $\chi$ is of first order. Thus, $\dot{\theta}$
is of first order and we can drop the last term in equation
(\ref{enteq2}) as it is of second order.
\item The perturbative expression for the expansion parameter is
\begin{equation}
\Theta \, = \, 3H + \epsilon F + \epsilon^{2}G + {\rm higher\, order\, terms} \, .
\end{equation}
We will only need to consider the zero'th order term since effects of the other
terms would be of higher order in perturbation theory.
\end{itemize}

Making use of the above points, the equation (\ref{enteq}) at leading order becomes
\begin{equation}
\ddot{s}_{a} + 3H\dot{s}_{a} + g^{2}\phi^{2}s_{a} \, = \, 0 \, .
\end{equation}
If we do a conformal transformation
\bea
a s_{a} \, &\equiv& \, S_{a} \nonumber \\
 a \phi \, &\equiv& \, \varphi
\eea
and work with conformal time instead of physical time we get
\begin{equation} \label{enteq3}
S_{a}'' + g^{2}\varphi^{2}S_{a} \, = \, 0 \, ,
\end{equation}
which has the same structure as the equations (\ref{Xeq1}, \ref{Xeq2}) for the $\chi$
background field. From the discussion of the solutions of this
equation in the early section of this article it thus follows
that
\begin{equation}
S_{a} \, = \, B_{1}u_{1a} + B_{2}u_{2a} \, ,
\end{equation}
where $u_{1a}$ is exponentially growing and $u_{2a}$ is exponentially
damped. Considering only the growing mode and remembering the same
form for $u_{1a}$ as we used in equation (\ref{growing}), we find
\begin{equation} \label{entresult}
S_{a} \, = \,  H k^{-3/2}\exp(\mu x)P_{1a}(x, k) \, ,
\end{equation}
where $P_{1a}$ indicate periodic functions of $x$ with
unit amplitude, and we hence conclude that the entropy
component is exponentially growing due to
parametric resonance at the beginning of preheating. This is one
of the main results of our paper.

We are interested in the process of conversion of the entropy
fluctuation into a curvature fluctuation. This process happens
continuously througout the preheating phase.
In the rest of this section we will study the evolution of curvature
covector due to this conversion process.

As is well known, entropy fluctuations can seed a growing
curvature fluctuation mode on super-Hubble scales. In the
linear approximation which we use (and working under the small
angle $\theta$ assumption) the induced curvature
fluctuation $\zeta^{ent}_k$ is given by
\be
\zeta^{ent}_k \, \simeq \, \frac{H}{\dot{\varphi}^2} \dot{\chi} S_k \, ,
\ee
where $S_k$ is the Fourier space entropy fluctuation determined
above in (\ref{entresult}). Making use of (\ref{entresult}),
inserting the result for $\chi$ given in
(\ref{background}), and taking care of the
change in the temporal variable from $x$ to $\eta$ we find
\be
\zeta^{ent}_k \, \simeq \, \sqrt{\pi \lambda} \frac{H^3 \varphi}{{\dot{\varphi}}^2}
e^{2 \mu x} k^{-3/2} P(x, k) \,
\ee
where $P$ is a periodic function of unit amplitude. This clearly
shows the exponential growth of the induced curvature fluctuations.
A derivation of this result from first principles making use of the
covariant formalism is given in the Appendix. Note that even though $\mu_k < 1$,
the time scale of the exponential instability is (while long
compared to the oscillation time) short compared
to the Hubble expansion time, the time scale relevant to the
conversion of entropy fluctuations to adiabatic ones.

We can evaluate the power spectrum for the
curvature perturbation given by the entropy mode from at horizon crossing:
\begin{equation}
\label{ps_inflation}
P_{k}^{ent}=\frac{k^3}{2 \pi^2} | \zeta _{k}^{ent} |^{2} \sim \lambda \frac{H^6}{\dot{\varphi}^4}\varphi ^{2} e^{4 \mu x}\,.
\end{equation}
Hence, we conclude that the entropy perturbations give a scale invariant contribution
to the power spectrum of the curvature perturbation from inflation, but with
an amplitude which is exponentially increasing.

We have derived our result in a simple two field inflation model, the
conclusions will carry over to other multi-field models. Our work suggests
that in any inflationary model in which the inflaton
satisfies the massless preheating condition, then if low mass
entropy fields are present which couple to the inflaton, then
parametric resonance of the entropy perturbation indeed happens. Due
to the conversion process of entropy perturbation into adiabatic perturbation
(as studied in the context of coupled scalar fields
in \cite{Gordon:2000hv}), parametric resonance of entropy perturbation
may lead to a rapidly growing adiabatic mode which could have a large impact on
the spectrum of curvature perturbation we observe today. The spectrum
will remain approximately scale-invariant, but there is the danger that
the exponential growth will cause the fluctuations to become non-linear
(which would rule out the model). To see whether this is a serious
concern, we must however first consider back-reaction issues. Back-reaction
might cut off the instability before the induced curvature fluctuations become
too large. However, below we find that at least the back-reaction effects
which we consider are not strong enough to shut off the resonance in time
for the induced curvature fluctuations to remain small enough.

\subsection{Back-reaction}

In the previous sections we considered the parametric resonance of entropy
perturbations during  preheating neglecting any back-reaction effects. However,
the exponential instability of the entropy field leads to an exponential creation of
$\chi$ particles that are expected to back react in the background. The study of
back-reaction is important, since the cumulative effect of the creation of particle eventually
becomes important  affecting the resonance and even terminating preheating, as already
noticed in \cite{Zibin,Greene:1997fu, KLS2} (see also \cite{KTk} for earlier
numerical work).

We will consider two back-reaction effects that can affect preheating in the
$g^2/\lambda =2$ case. Other back-reaction effects and parameters choices were
studied in \cite{Zibin,Greene:1997fu}. The first effect is the back reaction of the parametrically
amplified $\chi$ on the evolution of the inflaton background. If the force induced
by $\chi$ is larger than the force present in the absence of $\chi$, then the
condition for massless preheating will no longer be satisfied and the broad
parametric resonance will terminate. This will happen when
\begin{equation} \label{cond3}
g^2 \langle \chi ^2 \rangle \sim \lambda \phi ^2 \, .
\end{equation}
Using $\langle \chi ^2 \rangle = \chi _{eff}^2$ from (\ref{background}) and setting
$\phi=\phi_{end}$, this condition implies that
\begin{equation}
\label{efoldings_backreaction}
e^{2 \mu \Delta x} \simeq \lambda ^{-1}\,.
\end{equation}
This gives us the time interval before this back-reaction effect  becomes important.
We can use this result to evaluate the power spectrum of the curvature perturbations
induced by entropy modes at the time that the resonance shuts off by this
back-reaaction effect. Using (\ref{ps_inflation}) and (\ref{efoldings_backreaction}),
the power spectrum at this time is given by:
\begin{equation}
P_{k} ^{ent} \sim \lambda^{-1} \frac{H^6}{\dot{\varphi}^4} \, \varphi ^2\,.
\end{equation}
Since at the end of inflation $\dot{\varphi}^2 = V$ and $\varphi=\varphi_{end}$, we
can estimate the power spectrum for the curvature perturbations from the super-Hubble
amplified entropy perturbations as:
\begin{equation} \label{concl}
P_{k} ^{ent} \sim \frac{1}{5}\,.
\end{equation}
This is considerably larger than the observed values, exceeding by many orders of
magnitude the COBE normalization measurement \cite{Bunn:1996py}. We thus conclude
that the parametric amplification of entropy perturbations can lead to a serious problem for
models like the one we consider, unless other effects are found which shut off the resonance
earlier.

We can also consider the back-reaction of the produced $\chi$ particles on the Friedmann
equation. We find that demanding that the induced $\chi$ terms remain sub-dominant leads
to precisely the condition (\ref{cond3}).

The second effect considered in this paper is the influence of the
produced $\delta \phi$ particles on the $\delta \chi$ resonance. If the creation of $\phi$
particles is large enough, increasing significantly the effective mass of the
$\delta \chi$ field, this could damp or even
stop the resonance of the $\chi$ field. Thus we need to know if
\begin{equation}
V_{,\chi} < \, g^2 \chi \langle \delta \phi ^2 \rangle \,\,
\Rightarrow \,\,   \phi^2 < \langle \delta \phi ^2 \rangle \,,
\end{equation}
at some point during preheating,
altering the effective time dependent mass and consequently the $\chi$ resonance.
However, we can see from the equation for the eigenmodes $\phi_{k}(t)$ during preheating:
\begin{equation}
\ddot{\delta \phi_{k}}  + 3H\dot{\delta \phi_{k}} + \left( \frac{k^2}{a^2} +
3 \lambda \delta \phi^2 \right) \delta \phi_{k} \, = \, 0\,,
\end{equation}
that for $\delta \phi _{k}$ the resonance is always narrow, since it is equivalent to the
case of parametric resonance with $g^2 / \lambda =3$. This leads to a very small
characteristic exponent $\mu$  \cite{Greene:1997fu} and a very inefficient creation
of $\phi$ particles. This effect will always be less important than the parametric resonance
for the $\chi$ field, that is very broad and has a large characteristic exponent.
Thus, the second back-reaction effect studied in this paragraph does not have the potential
of shutting off the resonant amplification of entropy fluctuations early, and does not change
the conclusion from (\ref{concl}) that the curvature perturbations are too large when
the resonance is finally shut off.

We have to stress, though, that the full theory of back reaction and reescatering during
preheating is not fully developed. However, this result represents an advance with
respect to previous investigations since the covariant formalism allows for a full non-linear
analysis including metric fluctuations (although we here considered only the linear limit).

\section{Conclusions}

We have considered the preheating of entropy fluctuations in a two field model in
which an inflaton field with vanishing bare mass is coupled to a massless entropy
field. In the absence of metric fluctuations, it is known that in this model there is
efficient preheating (``massless preheating''). We find, using a covariant formulation
of the theory of cosmological fluctuations which in principle can be extended to a
full nonlinear analysis, that the entropy fluctuations experience a period of broad
parametric resonance. At quadratic order in the amplitude of fluctuations, the entropy
modes seed a growing curvature fluctuation. Hence, we find a curvature fluctuation
mode which is growing exponentially during the preheating phase. In
agreement with previous studies \cite{Zibin} we find that back-reaction effects are
too weak to shut off the resonance before the power spectrum of the
induced curvature fluctuations has reached an amplitude close to $1$, i.e. many
orders of magnitude larger than the observational value. Hence we
see that models of the type we consider here are phenomenologically
ruled out, unless there are back-reaction effects not considered here which
manage to truncate the resonance earlier than the ones we have studied.

Our results confirm those of earlier studies \cite{BV, FB2} which were reached using
different methods. The advantage of the formalism used here is that it is in
principle extendible to the non-linear regime, although here we use a linear
truncation of the method. 
Hence, our paper resolves a controversy about the validity of previous works.

A  significant limitation of our results is that they only apply to models with
a massless inflaton. Such models may be motivated by conformal symmetry
arguments. We do not regard the restriction to massless entropy fields
to be a significant limitation, the reason being that the conclusions
go through as long as the $\chi$ mass is small compared to $H$, and
there are lots of fields with masses smaller than the value of $H$ during
inflation.

\section{Acknowledgements}

We thank  A. Abolhasani, M. Oltean, L. Perreault Levasseur, M. Namjoo, M. Sasaki,
T. Suyama and Y. Wang for useful discussions. This work was supported in
part by a NSERC Discovery grant and by funds from the Canada
Research Chair program (RB).

\section{Appendix}

In this appendix we derive the growth of induced curvature fluctuations
from first principles, making use of the covariant formalism.

In the long wavelength
limit, the growth of $\zeta$ is given by \cite{Langlois:2005qp,Langlois:2006vv}
\begin{equation} \label{zetaeq2}
\dot{\zeta}_{a} \, \equiv \, \mathcal{L}_{u}\zeta_{a} \,
= \, \frac{2}{3}\frac{\Theta}{\dot{\sigma}^{2}}V_{,s}S_{a} \, .
\end{equation}
Note that the induced growing mode of $\zeta$ is quadratic in the magnitude
of fluctuations.

The equation (\ref{zetaeq2}) can be understood in the following way:
For the Lie derivative of curvature covector we have
\begin{equation}
\mathcal{L}_{u}\zeta_{a} \, = \, u^{c}\nabla_{c}\zeta_{a} + \zeta_{c}\nabla_{a}u^{c} \, .
\end{equation}
Considering $u^{c} = \{1/a,0,0,0\}$ we get
\begin{equation}
\mathcal{L}_{u}\zeta_{a} \, = \, \frac{1}{a}\partial_{t}\zeta_{a} \, .
\end{equation}
Then one can find the relation between the curvature covector and the conventional
coordinate based curvature perturbation as follows \cite{Langlois:2010vx}
(up to first order)
\begin{equation}
\zeta_{i} \, = \, \partial_{i}\zeta \, .
\end{equation}
Note that the quantity $\zeta$ we introduced here is the
same as the conventional curvature perturbation in the large scale limit which
we are interested in (see the discussion in \cite{Langlois:2010vx}).
The same relation holds for $S_{a}$, and considering $\partial_{i}\rightarrow ik$
in Fourier space leads to
\begin{equation}
\label{zeta_dot}
\frac{1}{a}\partial_{t}\zeta \, = \, \frac{\sqrt{2}H}{\dot{\sigma}^{2}}V_{,s}  H \, k^{-3/2} \exp(\mu x)P_{1}(x) \, .
\end{equation}

Up to first order in perturbation theory
\begin{equation}
\dot{\sigma}^{2} \, \simeq \, \dot{\phi}^{2}
\end{equation}
and
\begin{equation}
V_{,s} \, = \, -\lambda\phi^{3}\theta+g^{2}\phi^{2}\chi \, .
\end{equation}
Using dimensionless conformal time and again applying the conformal transformation
$\varphi=a\phi$ and the same for the preheat field $\chi$, we obtain for the
angle $\theta$
\begin{equation}
\theta \, = \, \frac{\frac{\chi'}{a}-\frac{a'}{a^{2}}\chi}{\frac{\varphi'}{a}-\frac{a'}{a^{2}}\varphi} \, .
\end{equation}
Therefore using equation (\ref{zetaeq2})  gives
\begin{equation}
\zeta ' =\frac{2 \sqrt{\pi}}{x} k^{-3/2} cos^{2} (x) e^{2\mu x} F(x) \, ,
\end{equation}
where
\begin{widetext}
\be
F(x)\equiv\lambda\frac{\mu x\sin^{2}(x)\cos(x)+x\cos^{2}(x)\sin(x)-\sin^{2}(x)\cos(x)}{(x\sin(x)+\cos(x))^{3}}+g^{2}\frac{\sin^{2}(x)}{(x\sin(x)+\cos(x))^{2}} \, ,
\ee
\end{widetext}
which clearly shows that
\begin{equation}
\zeta \, \propto \, \exp(2\mu x) \, .
\end{equation}
Hence, we conclude that parametric resonance of entropy perturbations
induces an exponentially growing curvature mode.

Thus, we see that taking into account the squeezing of the modes
and the solution (\ref{growing}), the curvature perturbation induced by
the entropy modes after inflation, using (\ref{zeta_dot}), is given by:
\begin{equation}
\zeta^{ent}_k \sim \sqrt{2 \pi \lambda} \, \frac{H^3}{\dot{\varphi}^{2}} \varphi \, k^{-3/2} e^{2 \mu x} \,,
\end{equation}
where we made the approximation of $V_{,s} \simeq \lambda \varphi ^{2} \chi _{eff}$
since $g^2/\lambda =2$, the same result obtained in the main part of the text
using the approximate treatment.

\end{document}